\documentclass[letterpaper]{article} 
\usepackage{aaai2026}  
\usepackage{times}  
\usepackage{helvet}  
\usepackage{courier}  
\usepackage[hyphens]{url}  
\usepackage{graphicx} 
\usepackage{natbib}  
\usepackage{caption} 
\usepackage{algorithm}
\usepackage{algorithmic}

\usepackage{amsmath}
\usepackage{amsfonts}
\usepackage{booktabs}
\usepackage{pifont}
\usepackage{newfloat}
\usepackage{listings}
\DeclareCaptionStyle{ruled}{labelfont=normalfont,labelsep=colon,strut=off} 
\floatstyle{ruled}
\newfloat{listing}{tb}{lst}{}
\floatname{listing}{Listing}
\title{AccKV: Towards Efficient Audio-Video LLMs Inference via Adaptive-Focusing and Cross-Calibration KV Cache Optimization}
\author{
    Zhonghua Jiang\textsuperscript{\rm 1}\equalcontrib,
    Kui Chen\textsuperscript{\rm 1}\equalcontrib, 
    Kunxi Li\textsuperscript{\rm 1}\equalcontrib, 
    Keting Yin\textsuperscript{\rm 1}, 
    Yiyun Zhou\textsuperscript{\rm 1},\\
    Zhaode Wang\textsuperscript{\rm 2}, 
    Chengfei Lv\textsuperscript{\rm 2},
    Shengyu Zhang\textsuperscript{\rm 1}\thanks{The corresponding author.}
}
\affiliations{
    \textsuperscript{\rm 1}Zhejiang University, 
    \textsuperscript{\rm 2}Alibaba Group\\
    \{jiangzhonghua, chen\_kui, kunxili, yinkt, yiyunzhou, sy\_zhang\}@zju.edu.cn \\
    {\{zhaode.wzd, chengfei.lcf\}@alibaba-inc.com}
}

\usepackage{bibentry}

\begin{document}

\maketitle

\begin{abstract}
Recent advancements in Audio-Video Large Language Models (AV-LLMs) have enhanced their capabilities in tasks like audio-visual question answering and multimodal dialog systems. 
Video and audio introduce an extended temporal dimension, resulting in a larger key-value (KV) cache compared to static image embedding. 
A naive optimization strategy is to selectively focus on and retain KV caches of audio or video based on task. However, in the experiment, we observed that the attention of AV-LLMs to various modalities in the high layers is not strictly dependent on the task. In higher layers, the attention of AV-LLMs shifts more towards the video modality. In addition, we also found that directly integrating temporal KV of audio and spatial-temporal KV of video may lead to information confusion and significant performance degradation of AV-LLMs. If audio and video are processed indiscriminately, it may also lead to excessive compression or reservation of a certain modality, thereby disrupting the alignment between modalities. 
To address these challenges, we propose AccKV, an Adaptive-Focusing and Cross-Calibration KV cache optimization framework designed specifically for efficient AV-LLMs inference. Our method is based on layer adaptive focusing technology, selectively focusing on key modalities according to the characteristics of different layers, and enhances the recognition of heavy hitter tokens through attention redistribution. In addition, we propose a Cross-Calibration technique that first integrates inefficient KV caches within the audio and video modalities, and then aligns low-priority modalities with high-priority modalities to selectively evict KV cache of low-priority modalities. 
The experimental results show that AccKV can significantly improve the computational efficiency of AV-LLMs while maintaining accuracy.
\end{abstract}

\section{Introduction}

In recent years, the rapid development of Audio-Video Large Language Models (AV-LLMs)~\cite{cheng2024videollama2,zhang2025videollama3,tang2024avicuna,chen2023vast,shu2023avllm} has significantly improved their performance in tasks such as video captioning~\cite{wang2019vatex,geng2024longvale,liu2025valor,xu2016msr}, audio-visual question answering~\cite{sun2024video,chen2020vggsound,wu2024star}, and multimodal dialog systems~\cite{team2023gemini,alamri2019avsd,chen2025taoavatar,wang2025personalized,yu2024gaussianTalker}. 
AV-LLMs, like VideoLLaMA2~\cite{cheng2024videollama2} and AVicuna~\cite{tang2024avicuna}, realize an efficient understanding of audio and video signals by integrating spatial-temporal modeling and cross-modal alignment technology. 
However, as the scale of the AV-LLMs expands, the memory usage problem of key-value (KV) cache in the inference process becomes increasingly prominent, especially for audio and video data, which naturally have long sequences and high redundancy characteristics, further exacerbating the computational and storage overhead. How to efficiently manage multimodal KV cache has become the core challenge in the actual deployment of AV-LLMs.

In order to speed up the reasoning process of Large Language Models (LLMs)~\cite{achiam2023gpt4,alexandre2023mistral,zhou2025cuff,qwen2,zhou2025revisiting}, several methods~\cite{shi2024keepcostdown,li2024survey,li2025madakv,jiang2025purekv,li2025flowmm} have been proposed to optimize the KV cache. The KV cache scheme based on dynamic attention filtering (such as H2O~\cite{zhang2023h2o}, SnapKV~\cite{li2024snapkv}) significantly reduces computing cost by maintaining a sparse set of important KV caches, but its design paradigm focuses on text scenarios and fails to adapt to heterogeneous redundancy patterns in multimodal data~\cite{li2024mvbench,alamri2019avsd,song2024milebench,maaz2023vcgpt}. LOOK-M~\cite{wan2024look} and FastV~\cite{chen2024fastv} formulate KV cache evaluation and merging strategies for the redundancy of visual tokens. However, due to AV-LLMs' unique attention pattern to audio and video and the unique spatial-temporal characteristics of audio and video data, the existing KV cache strategy cannot be directly applied to AV-LLMs.

Video and audio introduce extended temporal dimensions, leading to larger key-value caches compared to static image embeddings. A simple optimization strategy is to selectively focus on and retain the KV cache of audio or video according to specific task requirements. However, we observe that:
\begin{itemize}
\item \textbf{Attention Convergence:} In high layers, AV-LLMs' attention to each modality is not strictly dependent on specific tasks. Regardless of the task type (such as audio question and answer or video description), AV-LLMs tend to focus more on the video modality in higher layers.
\item \textbf{Heterogeneous modality merging conflict:} Directly merging the KV cache of audio with temporal characteristics and video with spatial-temporal characteristics may cause information confusion, thus significantly reducing AV-LLMs performance. This is because the redundancy characteristics and distribution of different modalities are significantly different. A simple direct merging will destroy the structure of modal information and cause confusion of feature distribution.
\item \textbf{Excessive compression/reservation trap:} If audio and video are processed indiscriminately, it may also lead to excessive compression or reservation of a certain modality, and destroy the balance and alignment between different modalities. In many real-world scenes, audio and visual events are synchronized (for example, voice and lip movements). AV-LLMs need to learn this synchronization to understand the content correctly.
\end{itemize}

To solve these problems, we propose AccKV, an Adaptive-Focusing and Cross-Calibration KV cache optimization framework. AccKV dynamically focuses on key modalities according to AV-LLMs' attention pattern in different layers through layer adaptive focusing. At the same time, Cross-Calibration technology is introduced to align the KV cache of low-priority and high-priority modalities after merging the redundancy within each modality, so as to avoid information confusion. In addition, existing methods usually rely on accumulated attention scores to identify important tokens. However, because the attention matrix is a lower triangular matrix, the early token is more likely to be selected as the key token, and the accumulated score of the most recent token can never be greater than the first token. AccKV uses attention redistribution to enhance the recognition of important tokens by reallocating attention weight.

Specifically, AccKV conducts an in-depth analysis of the attention patterns of audio and video modality tokens in different layers of AV-LLMs, dynamically calculates the modality priority of audio and video in each layer, and assigns differentiated weights accordingly, effectively focusing on key modalities. 
At the same time, we introduced an attention redistribution mechanism that improves the accuracy of identifying important tokens by readjusting the distribution of attention scores, assigning lower weights to attention scores related to early tokens and higher weights to attention scores related to more recent tokens. In addition, AccKV adopts Cross-Calibration technology, which first merges inefficient KV caches within each modality. This step reduces unnecessary information redundancy and effectively avoids information confusion caused by directly integrating different characteristic tokens. Subsequently, a cross-modal aligner based on cosine similarity was designed, using the K state of high-priority modalities as anchor points to perform threshold filtering on low-priority modalities, ensuring alignment between the two modalities.

The experimental results show that AccKV can significantly reduce the computational cost of AV-LLMs while maintaining accuracy. In summarize, this paper makes the following key contributions:  
\begin{itemize}
\item We observed a unique attention pattern in AV-LLMs and proposed a new Adaptive-Focusing and Cross-Calibration KV cache optimization framework based on it, named AccKV.
\item Based on AccKV framework, we have implemented a baseline that dynamically focuses on the key modality according to the attention patterns of AV-LLMs in different layers. At the same time, Cross-Calibration technology is introduced to compress the KV cache within each modality, and then align the KV cache of low-priority and high-priority modality. In addition, we use attention redistribution techniques to enhance the recognition of important tokens.
\item We conducted extensive experimental validation of the AccKV framework and tested it on two AV-LLMs (VideoLLaMA2~\cite{cheng2024videollama2} based on Qwen2-7B-Instruction~\cite{qwen2} language decoder and AVicuna~\cite{tang2024avicuna} based on Vicuna-7B-v1.5~\cite{zheng2023vicuna} language decoder). The experiment covered multiple datasets, and the results showed that AccKV can achieve efficient inference while maintaining performance in different scenarios, verifying its effectiveness and universality.
\end{itemize}

\begin{figure*}[htp]
\centering
\includegraphics[width=0.8\textwidth]{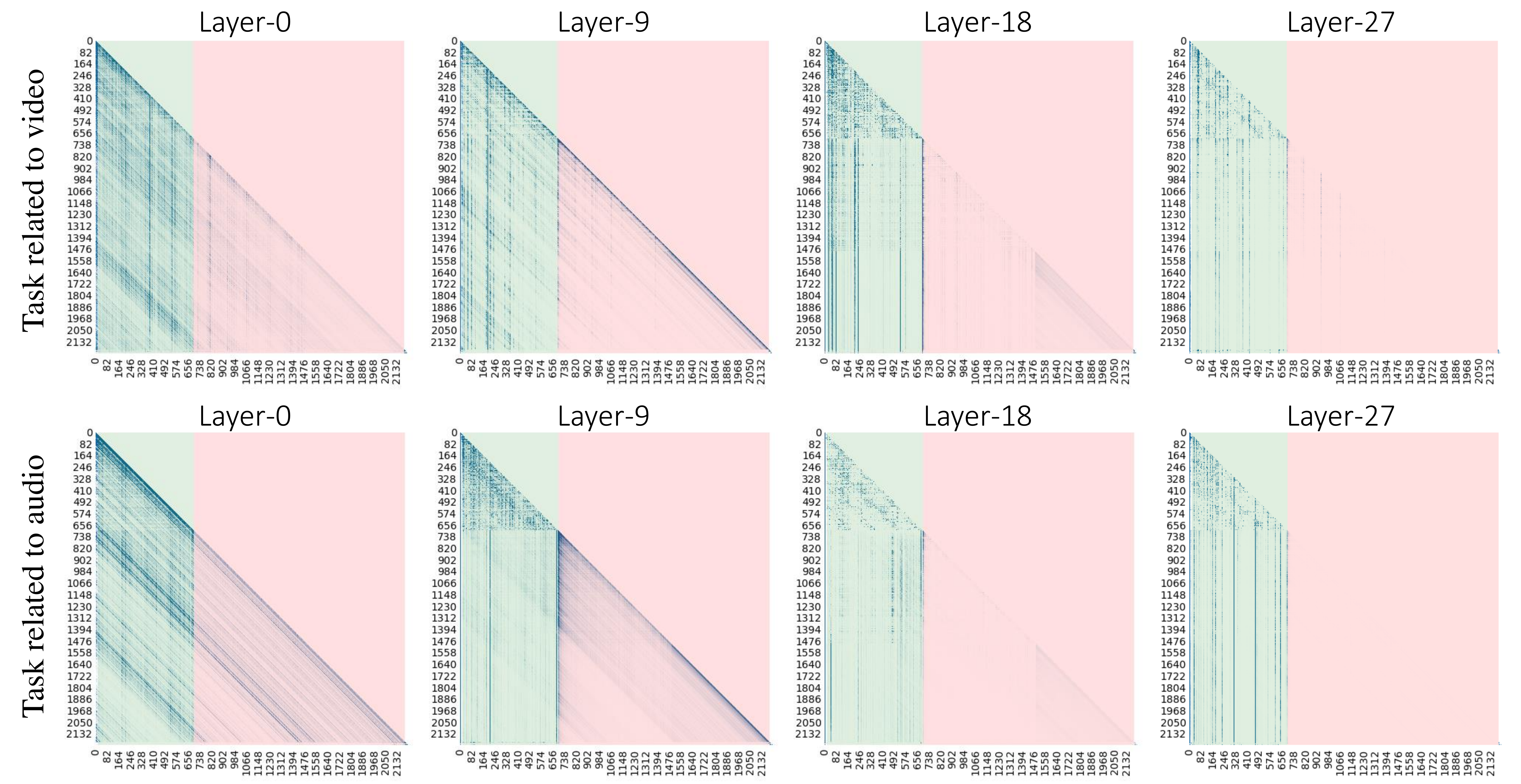}
  \caption{The attention distribution of VideoLLaMA2. Green and pink denote video and audio modalities, respectively. Blue represents attention score, and the darker the color, the higher the score. In lower layers, audio and video modalities each receive partial attention. In higher layers, the attention of AV-LLMs converge to video modality, regardless of the task type.}
  \label{fig:Observation1}
\end{figure*}

\section{Related work}

In LLMs, KV cache, as a key optimization technique, improves inference efficiency by caching previously calculated Key and Value vectors to avoid repeated calculations in the decoding stage. 
To address the growing computational demand~\cite{feng2023efficient,zhou2025disentangled,li2025mergenet,wang2025mitigating,jiang2025fedcfa}, H2O~\cite{zhang2023h2o} introduces a mechanism based on cumulative attention scores to identify and ignore relatively unimportant tokens by maintaining the top-k most representative KV pairs, thereby reducing computational and space complexity. However, H2O~\cite{zhang2023h2o} ignores the fact that the importance of tokens may change dynamically with the context. The Quest~\cite{tang2406quest} strategy adopts a page-level KV cache management method, using the inherent sparsity characteristics of the self-attention mechanism, using the Key state in each page to estimate its importance, and selecting the most important pages for precise calculation, thereby effectively reducing the computational load, but failing to reduce space complexity. The KIVI~\cite{liu2024kivi} strategy focuses on the quantization characteristics of Key and Value vectors, and effectively reduces memory usage by implementing channel-level quantization on Key and token-level quantization on Value, while maintaining model performance. FastGen~\cite{ge2023fastgen} further analyzes the functional differences of different attention heads, customizes specific compression strategies for each head, selects important tokens based on cumulative attention scores, and shows significant efficiency improvements in processing long sequences. Although the above strategies have shown powerful capabilities in plain text LLM scenarios, they do not take into account the redundant characteristics of multimodal tokens~\cite{team2023gemini}, which limits their performance in MLLMs~\cite{zhou2025cola,feng2024evolved,feng2025unified}.

In MLLMs, the optimization of KV cache requires customized strategies for the redundant characteristics of multimodal tokens. LOOK-M~\cite{wan2024look} analyzes the attention scores in the prefilling phase, prioritizes text tokens, and dynamically adjusts the retention strategy of KV cache in the decoding phase. It achieves efficient KV cache management by merging similar tokens. MustDrop~\cite{liu2024mustdrop} eliminates redundancy in the spatial dimension by merging adjacent tokens with high similarity and establishes a set of key tokens in the visual encoding phase, evaluates the importance of tokens through global and individual attention scores in the prefilling phase, and further compresses the KV cache in the decoding phase. PruneVid~\cite{huang2024prunevid} calculates the attention scores of question tokens and image tokens, selectively prunes video features related to question tokens, and minimizes video redundancy by merging spatial-temporal tags, achieving efficient KV cache management. However, these strategies are mainly targeted at the visual modality, and do not fully consider the unique attention patterns of the AV-LLMs for the audio-visual modality and the spatial-temporal alignment characteristics of the audio-visual modality. Therefore, they cannot be applied directly to AV-LLMs.

\begin{algorithm}[tb]
\caption{AccKV Algorithm.}
\label{alg:acckv1}
\begin{algorithmic}[1]
\REQUIRE Attention score matrix $A\in \mathbb{R}^{l\times l}$, KV Cache $K$ and $V$, the budget size of cache $k$, threshold $\tau$.
\ENSURE Updated KV cache.
\STATE Apply Attention Redistribution: $A_{i,j} \gets \frac{i+1}{l-j}*A_{i,j}, \ \  \forall  i,j \in [0,l-1]$
\STATE Calculate modal priority weights $W_v$ and $W_a$ using Eq. (3) to (5).
\STATE Adjust attention scores based on $W_v$ and $W_a$.
\STATE Select top-$k$ tokens with highest accumulated attention scores using Eq. (7) to (14).
\STATE Merge unimportant tokens within the same modality using Eq. (15) and (16).
\STATE Perform cross-modal alignment using cosine similarity as described in Eq. (17) to (20).
\STATE Evict the KV cache of low-priority modalities base on threshold $\tau$.
\end{algorithmic}
\end{algorithm}

\section{Method}
\subsection{Observation}

\subsubsection{Attention convergence.} In order to further analyze the attention patterns of AV-LLMs in different layers, we observed the attention distribution of each layer in detail. As shown in Figure \ref{fig:Observation1}, we found that in the lower layers, the audio and video modalities each received partial attention. However, at higher levels, the attention of AV-LLMs tends to converge to the video modality, regardless of the task type (such as audio Q\&A or video description). This indicates that AV-LLMs tend to pay more attention to video modalities in higher layers. Through this observation, we can better understand the internal working mechanism of AV-LLMs and guide the design of more effective KV cache management strategies.

In addition, we have verified this phenomenon through further experiments. The results in Table \ref{tab:Observation1} show that evicting the KV cache of video tokens in the high layer of the model results in significant performance degradation for both audio and video related tasks. In contrast, evicting the audio token from the KV cache has a relatively small impact on the performance of AV-LLMs. This proves the existence of attention convergence in AV-LLMs.

\begin{table}[htbp]
\centering
\setlength{\tabcolsep}{2mm}
\begin{tabular}{lrrr}
\toprule
 & Full Cache & Evict Audio & Evict Video \\
\midrule
Audio Task & 0.700490 & 0.691272 & 0.523281 \\
Video Task & 0.541667 & 0.525415 & 0.356390 \\
\bottomrule
\end{tabular}
\caption{The impact of evicting video and audio KV cache in higher layers on the performance of AV-LLMs.}
\label{tab:Observation1}
\end{table}

\subsubsection{Heterogeneous modality merging conflict.} 
In experiment, we observed that the LOOK-M~\cite{wan2024look} method performed poorly in audio and video scenes. After in-depth analysis, we believe that the main reason for this phenomenon is that LOOK-M~\cite{wan2024look} directly merges the audio KV cache with temporal characteristics with the video KV cache with spatial-temporal characteristics. This direct merging strategy fails to fully consider the significant differences in redundancy characteristics and distributions between different modalities, thereby disrupting the structure of modal information, leading to confusion in feature distributions, and ultimately significantly reducing model performance.

To verify this hypothesis, we designed and conducted a small-scale experiment. In this experiment, we attempted to merge the unimportant KV cache in the audio modality with the unimportant KV cache in the video modality. The results in Table \ref{tab:Observation2} indicate that this merging operation indeed leads to a significant decrease in the performance of AV-LLMs, thus verifying our hypothesis.

\begin{table}[htbp]
\centering
\setlength{\tabcolsep}{3mm}
\begin{tabular}{lrrr}
\toprule
& Full Cache & Evict & Merge \\
\midrule
Audio Task & 0.700490 & 0.651984 & 0.166571 \\
Video Task & 0.541677 & 0.474414 & 0.120523 \\
\bottomrule
\end{tabular}
\caption{The impact of heterogeneous modal merging on the performance of AV-LLMs. Evict: Evict the unimportant KV cache in audio modality and the unimportant KV cache in video modality. Merge: Merge the unimportant KV cache in the audio modality with the unimportant KV cache in the video modality.}
\label{tab:Observation2}
\end{table}

\subsubsection{Excessive compression/reservation trap.} 
If audio and video are processed indiscriminately, it may also lead to excessive compression or reservation of a certain modality, and destroy the balance and alignment between different modalities. In many real-world scenarios, audio and visual events are synchronized (such as speech and lip movements). AV-LLMs need to learn this synchronicity in order to correctly understand the content.

We designed an experiment to verify the importance of modal alignment: directly evict all audio KV cache or all video KV cache, and observe the changes in AV-LLMs performance. Table \ref{tab:Observation3} shows that even if all KV caches of modalities that are not directly related to the task are evicted, the performance of AV-LLMs will decrease. This result shows that the balance and alignment between modalities are critical to AV-LLMs performance, and excessive compression or retention of information of a certain modality will destroy this balance, thus affecting the AV-LLMs' ability to understand and process multimodal information.
\begin{table}[htbp]
\centering
\setlength{\tabcolsep}{1mm}
\begin{tabular}{lrrr}
\toprule
& Full Cache & Evict All Audio & Evict All Video \\
\midrule
Audio Task & 0.700490 & 0.018601 & 0.56708 \\
Video Task & 0.541667 & 0.490000 & 0.09015 \\
\bottomrule
\end{tabular}
\caption{The impact of evicting all audio KV cache or all video KV cache on AV-LLMs performance.}
\label{tab:Observation3}
\end{table}

\subsection{AccKV}
In this section, we propose AccKV, an Adaptive-Focusing and Cross-Calibration KV cache optimization framework. AccKV adaptively focuses on key modalities based on the attention patterns of AV-LLMs in different layers. AccKV introduces Cross-Calibration technology to compress redundancy within the modality, and then aligns the KV cache of low priority and high priority modalities to avoid information confusion. In addition, AccKV adopts attention redistribution to enhance the recognition of important tokens.

\begin{figure}[t]
\centering
\includegraphics[width=1.0\columnwidth]{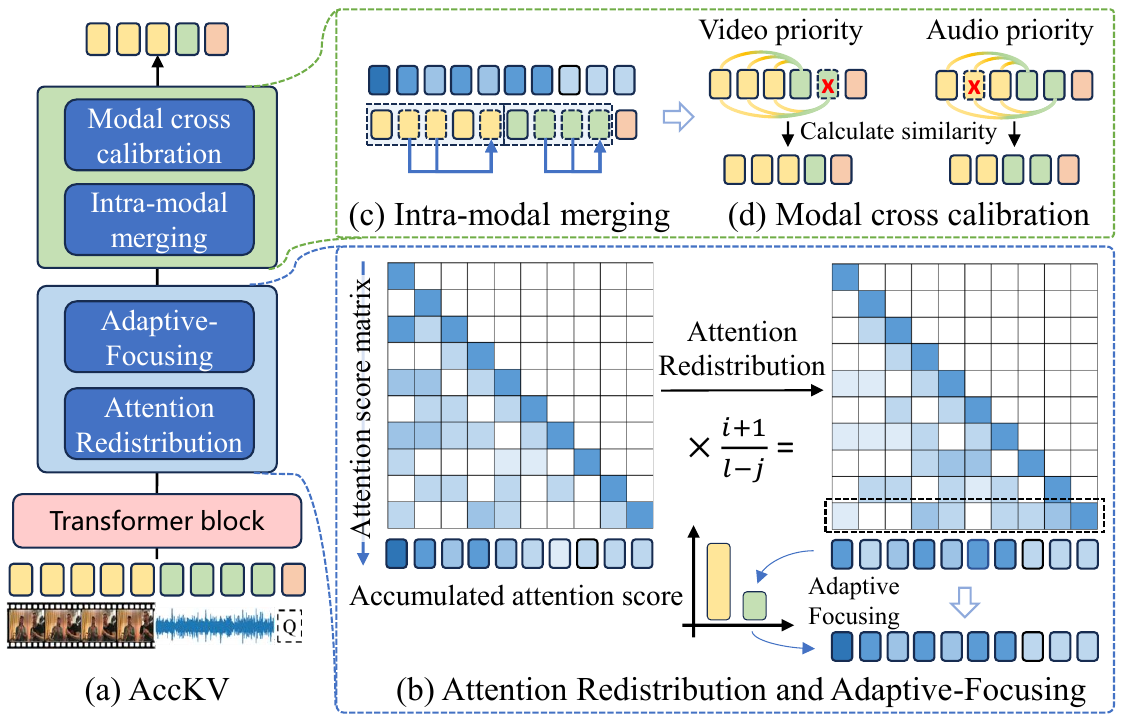}
  \caption{(a) Overview of AccKV. AccKV uses (b) attention redistribution to alleviate attention bias and adaptive focusing to identify important modalities, thereby avoiding misjudgment of KV cache importance. In addition, AccKV first (c) performs KV cache compression within each modality, and then (d) aligns low priority modalities with high priority modalities to maintain information structure integrity and reduce information confusion.}
  \label{fig:AcckV}
\end{figure}

\subsubsection{Attention Redistribution.} Most existing methods rely on accumulating attention scores to identify important tokens. However, due to the fact that the attention matrix is a lower triangular matrix, this mechanism tends to make early appearing tokens easier to identify as key tokens. To address this limitation, AccKV has introduced an attention reallocation that dynamically adjusts attention weights to more accurately identify the true important tokens.

Specifically, for a given attention matrix $A \in R^{l \times l}$, AccKV assigns lower weights to attention scores related to early tokens and higher weights to attention scores related to more recent tokens:
\begin{equation}
A_{i,j}=\frac{i+1}{l-j}*A_{i,j},
\end{equation}
\begin{equation}
A=softmax(\frac{QK^T}{\sqrt{d_k}}),
\end{equation}
where $l$ represents the length of the token sequence, $Q$ denotes the query matrix of the input tokens, $d_k$ is the dimension of K.

\subsubsection{Adaptive-Focusing.} Based on the observation results, we found that AV-LLMs exhibit different attention patterns in different layers. At the lower level, both audio and video modalities receive partial attention weights. It is worth noting that in higher levels, regardless of the type of task being performed, AV-LLMs tend to give more attention weight to video modalities. The AccKV framework dynamically calculates the modal priority of audio and video in different layers based on the attention patterns of audio and video modalities in AV-LLMs, and assigns differentiated weights accordingly, effectively focusing on key modalities.

Specifically, in each layer, AccKV calculates the average accumulated attention score of the text token at the end of the sequence with respect to the previously appearing video and audio tokens:
\begin{equation}
\bar{S}^v=\frac{\sum_{j=I_{0}^{v}}^{j<I_{1}^{v}}\sum_{i=I_{0}^{t}}^{i<l}A_{i,j}}{I_{1}^{v}-I_{0}^{v}},
\end{equation}
\begin{equation}
\bar{S}^a=\frac{\sum_{j=I_{0}^{a}}^{j<I_{1}^{a}}\sum_{i=I_{0}^{t}}^{i<l}A_{i,j}}{I_{1}^{a}-I_{0}^{a}},
\end{equation}
where the index range of the video token is $[I_0^v, I_1^v)$, the index range of the audio token is $[I_0^a, I_1^a)$, and the index range of the text token is $[I_0^t, l)$. Then we normalize $\bar{S}^v$ and $\bar{S}^a$ to obtain the modal priority of video and audio, which is used as the focus weight:
\begin{equation}
W^v=\frac{\bar{S}^v}{\bar{S}^v+\bar{S}^a}, \ \  W^a=\frac{\bar{S}^a}{\bar{S}^v+\bar{S}^a}.
\end{equation}
In order to focus AV-LLMs on the modality that this layer focuses on, AccKV weights the attention scores of video and audio:
\begin{equation}
A_{:,j} = \begin{cases}
 W^v*A_{:,j} & \text{ if } I_{0}^{v}\le j < I_{0}^{v}\\
 W^a*A_{:,j} & \text{ if } I_{0}^{a}\le j < I_{0}^{a}.
\end{cases}
\end{equation}

\subsubsection{Cross-Calibration.} The merging of KV cache can compress redundancy while preserving some information. 
However, directly merging the KV cache of audio with temporal characteristics and video with spatial-temporal characteristics may cause information confusion and affect model performance. 
Based on this observation, we propose merging inefficient KV caches within each modality. Firstly, we calculate the accumulated attention score of the text token at the end of the sequence for the previously appearing video and audio tokens:
\begin{equation}
S^v=\sum_{i=I_{0}^{t}}^{i<l}A_{i,j}, I_{0}^{v}\le j < I_{1}^{v}.
\end{equation}
Assuming we need to preserve the KV cache of k multimodal tokens, we will select these tokens based on the accumulated attention score. The index of the k multimodal tokens with the highest accumulated attention score is:
\begin{equation}
I_{top}=top_k([S^v,S^a], k).
\end{equation}
The number of reserved video and audio tokens are:
\begin{equation}
N^v=Min(Count(\{I|I \in I_{top}, I_{0}^{v}\le I < I_{1}^{v}\}), 1),
\end{equation}
\begin{equation}
N^a=Min(Count(\{I|I \in I_{top}, I_{0}^{a}\le I < I_{1}^{a}\}), 1).
\end{equation}
Then, we merge unimportant tokens within each modality. The index of the retained multimodal tokens are:
\begin{equation}
I_{top}^{v}=top_k([S^v], N^v),
\end{equation}
\begin{equation}
I_{top}^{a}=top_k([S^a], N^a).
\end{equation}
The index of the multimodal tokens that need to be merged is:
\begin{equation}
I_{merge}^v = \{I|I \notin I_{top}^v, I_{0}^v\le I < I_{1}^v \},
\end{equation}
\begin{equation}
I_{merge}^a = \{I|I \notin I_{top}^a, I_{0}^a\le I < I_{1}^a \}.
\end{equation}
Within each modality, we use the average merging method to handle the KV cache that needs to be merged:
\begin{equation}
\begin{split}
K_{merge}^v &= Avg(K[I_{merge}^v]), \ V_{merge}^v \\&= Avg(V[I_{merge}^v]),
\end{split}
\end{equation}
\begin{equation}
\begin{split}
K_{merge}^a &= Avg(K[I_{merge}^a]), \ V_{merge}^a \\&= Avg(V[I_{merge}^a]).
\end{split}
\end{equation}
Subsequently, we introduce a cross-modal alignment mechanism based on cosine similarity, using the K vectors of high-priority modalities as anchors to perform threshold filtering on low-priority modalities. Specifically, assuming that the video modality has a higher priority in a certain layer, i.e. $W^v>W^a$, we will calculate the cosine similarity of each K vector of the audio modality relative to all K vectors of the video modality:
\begin{equation}
\begin{split}
Sim&=Cosine\_Sim([K[I_{top}^{a}],K_{merge}^a], [K[I_{top}^{v}],K_{merge}^v]) \\& \in R^{(N^a+1) \times (N^v+1)}.
\end{split}
\end{equation}
Next, we average the cosine similarity between the K vectors of each audio token and the K vectors of all video tokens:
\begin{equation}
\overline{Sim}=Avg(Sim, dim=-1) \in R^{(N^a+1)}.
\end{equation}
We evict the KV cache of low-priority modalities with average similarity below the set threshold $\tau$, ensuring alignment with high-priority modalities. The last remaining KV cache is:
\begin{equation}
\begin{split}
K=[&K[I_{top}^{v}],K_{merge}^v,\\&filter(K[I_{top}^{a}],K_{merge}^a),K[I_0^t,l]],
\end{split}
\end{equation}
\begin{equation}
\begin{split}
V=[&V[I_{top}^{v}],V_{merge}^v,\\&filter(V[I_{top}^{a}],V_{merge}^a),V[I_0^t,l]].
\end{split}
\end{equation}
When the priority of the audio modality is higher, we adopt the same processing method.

\begin{table*}[ht]
\centering
\setlength{\tabcolsep}{3pt}
\small
\begin{tabular}{lrrrrrrrrrrr}
\toprule
& AS & AP & UA & OI & OS & AL & ST & AC & SC & CO & AVSD \\
\midrule
Full Cache  & 0.763425 & 0.782715 & 0.736952 & 0.80300   & 0.893333 & 0.737667 & 0.881910 & 0.515000 & 0.641925 & 0.58   & 0.479462 \\
\midrule
\multicolumn{12}{c}{20\% Cache Budget} \\
\midrule
H2O         & 0.678517 & 0.757552 & 0.624750 & 0.769631 & 0.856584 & 0.724232 & 0.821194 & 0.490000  & 0.610161 & 0.451667 & 0.352680  \\
SnapKV      & 0.711629 & 0.735932 & 0.642784 & 0.762555 & \underline{0.868453} & 0.721337 & 0.838955 & 0.485833 & 0.602885 & 0.469944 & \underline{0.424881} \\
LOOK-M      & 0.001667 & 0.001429 & 0.012052 & 0.017500 & 0.008333 & 0.008000  & 0.004345 & 0.000000   & 0.005000    & 0.035000  & 0.161588 \\
FastV       & 0.608172 & 0.643370 & 0.674452 & 0.629312 & 0.800072 & 0.653714 & 0.796022 & 0.452917 & 0.567033 & 0.490263 & 0.422403 \\
AccKV ($\tau$=0.9) & \textbf{0.748626} & \textbf{0.76603}  & \underline{0.683374} & \textbf{0.866752} & 0.866752 & \underline{0.736667} & \underline{0.854456} &\textbf{ 0.506667} &\textbf{ 0.641558} & \textbf{0.550000}   & \textbf{0.425179} \\
AccKV ($\tau$=0.6) & \underline{0.743455} & \underline{0.760238} & \textbf{0.695836} & \underline{0.785690}  & \textbf{0.869406} & \textbf{0.737667} & \textbf{0.86098}  & \underline{0.503333} & \underline{0.632954} & \underline{0.538750} & 0.423846 \\
\midrule
\multicolumn{12}{c}{10\% Cache Budget} \\
\midrule
H2O         & 0.571982 & 0.585202 & 0.508345 & 0.550764 & 0.726125 & 0.601895 & 0.705956 & 0.358417 & 0.458830  & 0.458830 & 0.332905 \\
SnapKV      & \underline{0.644221} & 0.664302 & 0.556322 & 0.660142 & 0.731768 & 0.653572 & 0.789548 & 0.383639 & 0.516122 & 0.364548 & 0.325341 \\
LOOK-M      & 0.001667 & 0.001111 & 0.012052 & 0.000000  & 0.004167 & 0.000000 & 0.004345 & 0.000000 & 0.005000 & 0.025000 & 0.160928 \\
FastV       & 0.587647 & 0.659611 & \textbf{0.703651} & 0.640790  & \underline{0.790262} & 0.638909 & 0.804543 & 0.407239 & 0.546424 & 0.462627 & 0.338811 \\
AccKV ($\tau$=0.9) & 0.642198 & \textbf{0.681243} & \underline{0.642874} &\textbf{ 0.707762} & 0.785710  & \textbf{0.663565} & \underline{0.829384} & \textbf{0.432333} & \textbf{0.602918} & \underline{0.497167} & \textbf{0.390516} \\
AccKV ($\tau$=0.6) & \textbf{0.649873} & \underline{0.675981} & 0.631588 & \underline{0.694291} & \textbf{0.803389} & \underline{0.660920}  & \textbf{0.830027} & \underline{0.421917} &\underline{ 0.593929} & \textbf{0.501500}  & \underline{0.385812} \\
\bottomrule
\end{tabular}
\caption{VideoLLaMA2: Performance of eviction strategy on MVBench and AVSD. The best results are highlighted in bold. The second result is highlighted with an underline.}
\label{tab:main_tab1}
\end{table*}

\vspace{-2em}

\begin{table*}[ht]
\centering
\setlength{\tabcolsep}{3pt}
\small
\begin{tabular}{lrrrrrrrrrrr}
\toprule
& AS & AP & UA & OI & OS & AL & ST & AC & SC & CO & AVSD \\
\midrule
Full Cache  & 0.557046 & 0.409944 & 0.434683 & 0.632619 & 0.563191 & 0.43307 & 0.656051 & 0.236181 & 0.505397 & 0.356611 & 0.321997 \\
\midrule
\multicolumn{12}{c}{160 tokens Cache Budget} \\
\midrule
H2O         & 0.102427 & 0.142147 & 0.157986 & 0.203755 & 0.238484 & 0.021622 & 0.214895 & 0.055501 & 0.172033 & 0.006933 & 0.046019 \\
SnapKV      & 0.277837 & 0.390677 & 0.167643 & 0.489307 & 0.336405 & 0.005714 & 0.229411 & 0.182790 & 0.274575 & 0.109262 & 0.274924 \\
LOOK-M      & 0.069215 & 0.095429 & 0.161076 & 0.108192 & 0.202830 & 0.089147 & 0.246255 & 0.037798 & 0.108152 & 0.000625 & 0.019048 \\
FastV       & 0.290436 & 0.308278 & 0.164896 & 0.497083 & 0.424673 & 0.044266 & 0.267676 & 0.098631 & 0.312610 & 0.043869 & 0.303037 \\
AccKV ($\tau$=0.9) & \underline{0.560320} & \textbf{0.421473} & \underline{0.433202} & \underline{0.629119} & \underline{0.551913} & \textbf{0.447873} & \underline{0.677995} & \underline{0.215736} & \textbf{0.505354} & \textbf{0.338278} & \underline{0.319018} \\
AccKV ($\tau$=0.6) & \textbf{0.567582} & \underline{0.420181} & \textbf{0.435402} & \textbf{0.637786} & \textbf{0.55907} & \underline{0.445151} & \textbf{0.680395} & \textbf{0.240208} & \underline{0.498489} & \underline{0.334302} & \textbf{0.321268} \\
\midrule
\multicolumn{12}{c}{120 tokens Cache Budget} \\
\midrule
H2O         & 0.015563 & 0.005615 & 0.086318 & 0.027991 & 0.174468 & 0.027225 & 0.286768 & 0.036946 & 0.089592 & 0.001000 & 0.035033 \\
SnapKV      & 0.255297 & 0.341656 & 0.350181 & 0.350181 & 0.065491 & 0.065142 & 0.240582 & 0.146657 & 0.173491 & 0.033500 & 0.303612 \\
LOOK-M      & 0.010662 & 0.006688 & 0.051570 & 0.020160 & 0.168745 & 0.014839 & 0.238061 & 0.115274 & 0.056398 & 0.002000 & 0.213519 \\
FastV       & 0.230038 & 0.277104 & 0.164116 & 0.342501 & 0.097991 & 0.144145 & 0.247935 & 0.108969 & 0.180847 & 0.068095 & 0.213519 \\
AccKV ($\tau$=0.9) & \textbf{0.556088} & \textbf{0.421328} & \underline{0.421478} & \underline{0.634119} & \underline{0.549646} & \textbf{0.433463} & \underline{0.665016} & \textbf{0.230069} & \textbf{0.510921} & \underline{0.358382} & \underline{0.315302} \\
AccKV ($\tau$=0.6) & \underline{0.555433} & \underline{0.418847} & \textbf{0.433581} & \textbf{0.636119} & \textbf{0.561166} & \underline{0.428963} & \textbf{0.667789} & \underline{0.229931} & \underline{0.510109} & \textbf{0.360734} & \textbf{0.319977} \\
\bottomrule
\end{tabular}
\caption{AVicuna: Performance of eviction strategy on MVBench and AVSD. The best results are highlighted in bold. The second result is highlighted with an underline.}
\label{tab:main_tab2}
\vspace{-0.4cm}
\end{table*}

\section{Experiments}
\subsection{Setting}
MVBench~\cite{li2024mvbench} is a comprehensive and challenging multimodal video understanding benchmark that covers 20 complex video understanding tasks. In our experiment, we evaluated 10 tasks that included both audio and video, including: Action Sequence (AS), Action Prediction (AP), Unexpected Action (UA), Object Interaction (OI), Object Shuffle (OS), Action Localization (AL), Scene Transition (ST), Action Count (AC), State Change (SC) and Character Order (CO). In addition, to further validate the effectiveness of the algorithm, we also conducted experiments using the Audio Visual Scene Aware Dialogue (AVSD)~\cite{alamri2019avsd} dataset. The AVSD~\cite{alamri2019avsd} dataset focuses on dialog understanding and provides rich audio-visual information. We use ROUGE as the evaluation metric

To comprehensively verify the effectiveness of AccKV, we conducted extensive experiments on two AV-LLMs equipped with different language decoders: VideoLLaMA2~\cite{cheng2024videollama2} based on Qwen2-7B-Instruct~\cite{qwen2} language decoder and AVicuna~\cite{tang2024avicuna} based on Vicuna-7B-v1.5~\cite{zheng2023vicuna} language decoder. Through this approach, we are able to evaluate the performance of AccKV under different architectures and decoding mechanisms.
In the experiment, we compared four representative KV caching strategies. Among them, H2O~\cite{zhang2023h2o} and SnapKV~\cite{li2024snapkv} focus on KV cache optimization for LLMs. FastV~\cite{chen2024fastv} and LOOK-M~\cite{wan2024look} are specifically optimized for KV caching in pure visual scenes. We conducted experiments on a NVIDIA A100 with 40GB memory.

\begin{table}[t]
\setlength{\tabcolsep}{9pt}
    \centering
    \setlength{\tabcolsep}{5mm}
    \begin{tabular}{ccc|c}
    \toprule 
        A-R & A-F & C-C & AVSD   \\ 
        \midrule
        \ding{52} & \ding{52} & \ding{52} & \textbf{0.425179}  \\ 
        \ding{52} & \ding{52} & \ding{56} & 0.364630  \\ 
        \ding{52} & \ding{56} & \ding{56} & 0.342471 \\ 
        \ding{56} & \ding{56} & \ding{56} & 0.334883  \\ 
        \bottomrule[1.25pt]
    \end{tabular}
    \caption{Ablation study. A-R: Attention Redistribution, A-F: Adaptive-Focusing, C-C: Cross-Calibration. The best results are highlighted in \textbf{bold}.}
    \label{tab:ablation}
    \vspace{-0.35cm}
\end{table}

\subsection{Experiment Results}
In Tables \ref{tab:main_tab1} and \ref{tab:main_tab2}, we conducted a comprehensive comparative analysis of AccKV with various eviction strategies in audio and video understanding scenarios. The experimental results clearly demonstrate the effectiveness of AccKV in managing KV cache in memory constrained environments, while maintaining high performance in various tasks. Specifically, AccKV achieved up to 80\%-90\% reduction in memory usage, with only a slight decrease in performance compared to a fully cached solution. This demonstrates the ability of AccKV to significantly reduce memory usage with minimal performance cost.

Compared with baseline eviction methods, AccKV consistently outperforms other methods on most datasets. It is worth noting that AccKV exhibits significant superiority over text-based unimodal KV exclusion methods. This discovery reveals a key limitation of text unimodal KV eviction methods: they often overlook unique redundant patterns in multimodal data, resulting in incorrect KV cache eviction and erroneous response generation. In addition, AccKV also performed well compared to LOOK-M and FastV. LOOK-M, as a method specifically designed for multimodal scenes, significantly reduces model performance by directly merging audio KV cache with temporal characteristics and video KV cache with spatial-temporal characteristics, resulting in information confusion. FastV failed to fully consider the unique attention pattern of AV-LLMs in the high layers, and indiscriminately evicted the audio and video KV cache, which may lead to excessive compression and disruption of inter-modal alignment. In contrast, AccKV uses layer-adaptive focusing technology to dynamically adjust modal weights at each layer based on its unique attention pattern. At the same time, AccKV effectively avoids information confusion and ensures alignment between modalities through modal internal merging and cross calibration mechanisms. AccKV maintains the performance of AV-LLMs while compressing KV cache. This method performs well in complex audio and video modal scenes, verifying its effectiveness and universality in multimodal scenes.

\subsection{Influence of Various Cache Budgets}
To evaluate the effectiveness of AccKV under different cache budget conditions, we conducted experiments on VideoLLaMA2 and AVicuna. The experimental results are presented in Tables \ref{tab:main_tab1} and \ref{tab:main_tab2}, respectively. As the cache budget decreases, the performance of other KV cache eviction strategies has declined significantly. In contrast, AccKV has demonstrated its outstanding robustness and efficiency. Specifically, under extremely strict memory constraints, such as VideoLLaMA2 retaining only 10\% of its cache budget and AVicuna retaining only 120 tokens of KV cache, AccKV can still maintain stable performance. This indicates that AccKV can accurately identify and retain key information in KV cache, thereby minimizing the risk of information loss while significantly reducing memory usage.

\subsection{Ablation Study}
To evaluate the individual contributions of each component in our proposed AccKV framework, we conducted an ablation study. Specifically, we examined the impact of three key modules: Attention Redistribution (A-R), Adaptive-Focusing (A-F), and Cross-Calibration (C-C). The results are summarized in Table \ref{tab:ablation}.
The result shows the performance of the model on the AVSD dataset when different combinations of these modules are enabled or disabled. When all three modules (A-R, A-F, and C-C) are active, the model achieves the highest performance.

These results highlight the importance of each module of AccKV. Attention Redistribution helps in more accurately identifying important tokens by adjusting attention weights dynamically. Adaptive-Focusing ensures that the model can dynamically focus on key modalities based on the layer's attention patterns. Cross-Calibration integrates redundant information within each modality and aligns low-priority and high-priority modalities effectively, avoiding information confusion. Together, these modules work synergistically to improve the model's efficiency and performance.

\section{Conclusion}
This study first observed some unique phenomena in AV-LLMs that are crucial for improving model inference efficiency. Specifically, we found attention convergence phenomenon, where in higher layers, AV-LLMs' attention to different modalities does not strictly depend on specific tasks. Regardless of the task type, AV-LLMs tend to focus more on the video modality in higher layers. At the same time, directly merging audio KV cache with temporal characteristics and video KV cache with spatial-temporal characteristics may lead to information confusion, significantly reducing the performance of AV-LLMs. If audio and video are processed indiscriminately, it may lead to over compression or reservation of a certain modality, and destroy the balance and synchronization between different modalities. Based on the above findings, we propose AccKV, which is specifically designed to improve the computational efficiency in the inference process of AV-LLMs. AccKV dynamically selects the key modality based on the attention patterns of different layers of AV-LLMs through Adaptive-Focusing technology, and uses Cross-Calibration technology to integrate redundant information internally before aligning low-priority and high-priority modalities to avoid information confusion. In addition, AccKV introduces an attention reallocation mechanism to more accurately identify important tokens by adjusting attention weights. The experimental results show that AccKV can significantly reduce the computational cost of AV-LLMs while maintaining model accuracy. This verifies the effectiveness and universality of AccKV.

\section{Acknowledgments}
This work was supported by the Key Research and Development Program of Zhejiang Province (No. 2025C01026), the National Natural Science Foundation of China (No. 62402429, U24A20326, 62441236), the Ningbo Yongjiang Talent Introduction Programme (2023A-397-G). Additionally, this work was partially supported by Young Elite Scientists Sponsorship Program by CAST (2024QNRC001). The author gratefully acknowledges the support of Zhejiang University Education Foundation Qizhen Scholar Foundation.

\bibliography{aaai2026}

\newpage
\section{Appendix}
\setcounter{table}{0}
\setcounter{figure}{0}
\renewcommand{\thetable}{\Roman{table}}
\renewcommand{\thefigure}{\Roman{figure}}

\newcommand{\vpara}[1]{\vspace{0.07in}\noindent\textbf{#1 }}
\setcounter{equation}{20} 

\subsection{Intuitive explanation of each module of AccKV and pseudocode of AccKV algorithm process}

We have added more intuitive explanations for each module of AccKV, 
and provided detailed pseudocode of the AccKV algorithm process to help readers better understand the overall working mechanism of the AccKV algorithm.

\vpara{Attention redistribution} explicitly compensates for the structural bias of the lower triangular matrix, alleviating the importance estimation bias of "tokens appearing earlier are more important", making the accumulated attention score more consistent with the true importance of tokens, and providing unbiased importance estimation for KV Cache optimization.

\vpara{Adaptive Focusing:} We observed that AV-LLMs have different attention patterns for each modality at different layers. Therefore, the layer adaptive focusing mechanism can dynamically identify the modal priority of each layer and allocate matching attention weights to the corresponding modality based on this, effectively focusing on key modalities.

\vpara{Cross-Calibration:} Due to the differences in the distribution characteristics of tokens in different modalities, direct cross modal merging can easily lead to information confusion (such as LOOK-M). Therefore, AccKV first merges low importance tokens within the modalities. Subsequently, in order to maintain alignment between modalities, AccKV introduced a token compression mechanism that guides low priority modalities with high priority modalities, i.e., based on the similarity evaluation with high priority modalities, the KV cache of low priority tokens with lower correlation is removed.

\begin{figure}[t]
\centering
\includegraphics[width=0.48\textwidth]{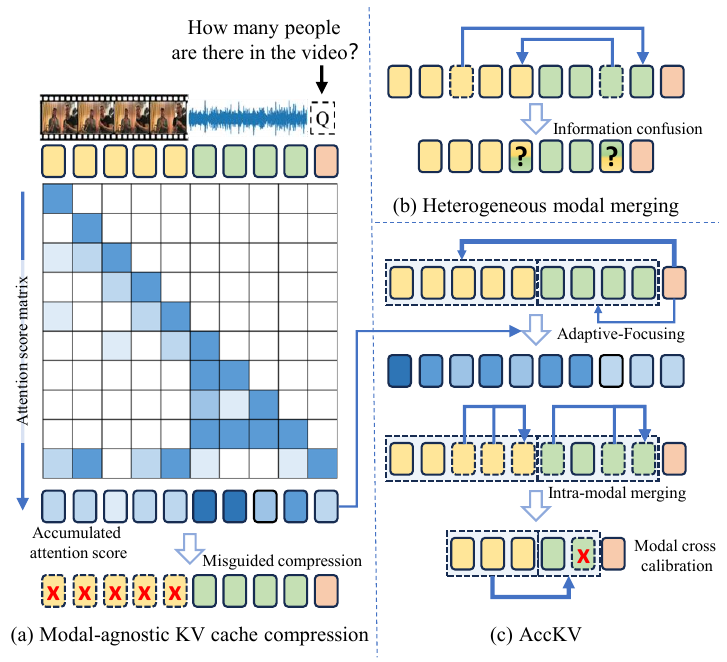}
\caption{(a) Modal-agnostic KV cache compression methods misidentify high-attention tokens within a single modality as globally important, as cumulative attention scores are confounded by intra-modal self-attention patterns. (b) Direct merging of heterogeneous modalities' KV caches disrupts intrinsic information structures and introduces cross-modal interference. (c) Our AccKV adopts adaptive focusing to identify important modes, thereby avoiding misjudgment of KV cache importance. It first performs KVcache compression within each modality, and then aligns low priority modalities with high priority modalities to maintain structural integrity and reduce information confusion.}
\label{fig:intro}
\end{figure}

\subsection{Analysis and experimentation of computational cost}

\vpara{The computational cost of the original attention mechanism.} The KV cache optimization algorithm mainly optimizes the decoding stage of AV-LLMs. As shown in Eq. (6) of the paper, for an input with a sequence length of $l$, when generating the next token, the attention mechanism mainly includes the following two calculation steps:

Attention score calculation: Calculate the attention score by multiplying the current query $q_l$ with the key matrix KT, and perform scaling and softmax operations: 
\begin{equation}
A_{l,:}=softmax(\frac{q_{l}K^T}{\sqrt{d_k}})\in\mathbb{R}^{1\times l},
\end{equation}
The computational cost of this step is $3l\cdot d_k$.

Therefore, the total computational cost of generating a token is $4l \dot d_k$. The total computational cost for generating n tokens is:
\begin{equation}
\begin{split}
    T_{raw}(l,n) &= {\textstyle \sum_{i=0}^{n-1}4\cdot(l+n)\cdot d_k} \\
    &=4nl\cdot d_k + 2n(n-1)d_k.
\end{split}
\end{equation}

\vpara{The computational cost of AccKV.} We considered retaining a 10\% KV cache and conducted a detailed analysis of the computational cost of the AccKV:

Therefore, the total computational cost of AccKV is:
\begin{equation}
\begin{split}
    T_{AccKV}(l,n) = &(0.5+0.0025d_k)l^2+llogl \\
    &+(2.5+0.9d_k+0.4nd_k)l \\
    &+2n(n-1)d_k.
\end{split}
\end{equation}
The computational cost saved by using AccKV is
\begin{equation}
\begin{split}
    T_{raw}(l,n)-T_{AccKV}(l,n) = &3.6nd_k-(0.5+0.0025d_k)l \\
    &-logl-2.5-0.9d_k.
\end{split}
\end{equation}

As the number of generated tokens increases, AccKV can significantly reduce computational cost, demonstrating its advantage in long sequence generation tasks.

As shown in Table \ref{tab:delay}, we measured the delay of AV-LLMs in the prefill and decoding stages through experiments, and verified the effectiveness of AccKV in improving the decoding speed of the model.

\begin{table*}[t]
\centering
\begin{tabular}{c|ccccc}
\toprule[1.25pt]
       & prefill & \multicolumn{3}{c}{decode} & total \\
\midrule
       & 2210    & 100     & 500    & 1000    & 3210  \\
\midrule
Origin & 181.17 & 3502.89 & 16302.19 & 32350.80 & 32531.97 \\
AccKV & 501.67 & 3157.31 & 15604.95 & 31406.54 & 31908.21   \\
$\Delta $ & -320.50 & 345.58 & 697.25 & 944.25 & 623.75  \\
\bottomrule[1.25pt]
\end{tabular}
\caption{Delay (ms) of VideoLLaMA2 in the prefilling and decoding stages.}
\label{tab:delay}
\end{table*}

\begin{table*}[t]
\centering
\setlength{\tabcolsep}{1.2mm}
\begin{tabular}{c|ccccccccccc}
\toprule[1.25pt]
$\tau$ & 0.01 & 0.1 & 0.2 & 0.3 & 0.4 & 0.5 & 0.6 & 0.7 & 0.8 & 0.9 & 1.0 \\
\midrule
AccKV & 396103 & 0.39611 & 0.395419 & 0.398557 & 0.401943 & 0.412595 & 0.423846 & 0.423845 & 0.423602 & 0.425179   & 0.413498 \\
\bottomrule[1.25pt]
\end{tabular}
\caption{Analysis of hyperparameter $\tau$.}
\label{tab:hyperparameter}
\end{table*}

\begin{table*}[thp]
\centering
\setlength{\tabcolsep}{1.2mm}
\begin{tabular}{c|ccc}
\toprule[1.25pt]
 & Average Merge & Attention score weighted merging & Normalized attention score weighted merging \\
 \midrule
AccKV & 0.425179 & 0.373752 & 0.353149 \\
\bottomrule[1.25pt]
\end{tabular}
\caption{Different KV cache merging methods.}
\label{tab:merging}
\end{table*}

\subsection{Further explanation and proof of Attention Redistribution}
We have supplemented the intuitive motivation, detailed explanation, and effectiveness proof of the Attention Redistribution mechanism in the paper to further elucidate the value and role of this module.

In autoregressive decoding, the attention score matrix of LLMs exhibits a lower triangular structure (i.e., only the main diagonal and its lower left elements are non-zero), which leads to traditional importance estimation methods based on cumulative attention scores tending to overestimate the importance of the early appearing tokens. In fact, without correction, the cumulative attention score of the last token cannot exceed that of the first token, which clearly does not conform to the distribution of actual semantic importance.

The Attention Redistribution module is introduced in AccKV as an explicit compensation mechanism for this positional bias, and its core idea includes the following two aspects:

\begin{itemize}
    \item Row weighting (Query side compensation): For Query token i located at the end of the sequence, it can "see" more Key tokens (i.e. more non-zero elements in the row), and theoretically requires more attention. Therefore, we assign higher weights to these later query rows.
    \item Column normalization (Key side compensation): For Key token $j$ located earlier in the sequence, it naturally accumulates higher scores in the original cumulative attention score due to its attention by all subsequent queries (i.e., the column has more non-zero elements). We suppress the "exposure advantage" caused by the position being earlier by normalizing in the column direction, thereby avoiding a systematic overestimation of the importance of early keys.
\end{itemize}

\subsubsection{Proof: Attention Redistribution alleviates positional bias}
Consider a situation where all tokens have the same importance. Let the sequence length be $l$ and the token index be $i, j \in [0, l-1]$. The attention matrix $A \in R ^ {l × l}$ is a lower triangular matrix that satisfies the causal constraint:
\begin{equation}
A_{i,j}=\begin{cases}
 \frac{1}{i+1} & \text{ if } j \le i \\
 0 & \text{ otherwise }
\end{cases}.
\end{equation}

\vpara{The positional bias of the original cumulative attention score.} Sum A by column, and the cumulative attention score of the j-th token is
\begin{equation}
S_j^{raw} ={\textstyle \sum_{i=j}^{l}A_{i,j}}={\textstyle\sum_{i=j}^{l} \frac{1}{i+1}}.
\end{equation}
The sum is a partial sum of harmonic series, satisfying:
\begin{equation}
S_j^{raw} = H_l-H_j,
\end{equation}
where $H_n = {\textstyle \sum_{k=1}^{n}\frac{1}{k}}$ is the n-th harmonic number. The monotonicity of harmonic numbers ($H_n$ increases with the increase of $n$) indicates that $S_j^{raw}$ decreases with the increase of $j$, that is: 
\begin{equation}
S_0^{raw}>S_1^{raw}> ... >S_{l-1}^{raw}.
\end{equation}
This result contradicts the assumption that "all tokens have the same importance", indicating that the positional bias of the original cumulative attention score can mislead the judgment of token importance.

\vpara{Attention Redistribution alleviates positional bias.} According to the revised formula (8), the attention score matrix after attention redistribution is: 
\begin{equation}
A_{i,j}^{redist}=\frac{i+1}{l-j} \cdot A_{i,j}=\begin{cases}
 \frac{1}{l-j} & \text{ if } j \le i \\
 0 & \text{ otherwise }
\end{cases}.
\end{equation}

The cumulative attention score of the j-th token is:
\begin{equation}
S_j^{redist} ={\textstyle \sum_{i=j}^{l}A_{i,j}^{redist}}={\textstyle\sum_{i=j}^{l} \frac{1}{l-j}}.
\end{equation}
Since $\frac{1}{l-j}$is a constant and the number of summation terms is $(l-1) - j+1=l-j$, therefore: 
\begin{equation}
S_j^{redist} = (l-j) \cdot \frac{1}{l-j}=1.
\end{equation}
It can be concluded that:
\begin{equation}
S_j^{raw}=S_k^{raw}=1, \forall j,k\in [0,l-1].
\end{equation}
The cumulative attention scores of all tokens are equal, which is consistent with the assumption that "all tokens have the same importance". 

So Attention Redistribution explicitly compensates for the structural bias of the lower triangular matrix, making the accumulated attention score more consistent with the true importance of the token, providing unbiased importance estimation for KV Cache optimization.

\subsection{More experiments}
\vpara{Analysis of hyperparameter $\tau$.} The threshold $\tau$ is used to filter the KV cache of inefficient tokens during the modal alignment stage. Specifically, AccKV removes the KV cache of low priority modalities with an average similarity below the set threshold $\tau$, thereby achieving effective alignment with high priority modalities. As shown in Table \ref{tab:hyperparameter}, We conducted experiments at multiple $\tau$ values (from 0.1 to 1.0), and the results showed that modal alignment performed best within the range of 0.6 to 0.9, avoiding excessive retention of redundant information and preventing information loss caused by excessive compression. AccKV exhibited good accuracy and stability within this range.

\vpara{Different KV cache merging methods.} We compared different KV cache merging methods and the Table \ref{tab:merging} results showed that using the average merging strategy performed the best in terms of performance. Since average merging does not rely on the importance estimation of any specific token, it has stronger robustness to fluctuations in input data and can effectively alleviate errors introduced by outliers or noise.

\end{document}